# Impacts of propagating, frustrated and surface modes on radiative, electrical and thermal losses in nanoscale-gap thermophotovoltaic power generators

Michael P. Bernardi,[1,a)] Olivier Dupré,[2] Etienne Blandre,[2] Pierre-Olivier Chapuis,[2] Rodolphe Vaillon,[2,b)] and Mathieu Francoeur[1,c)]

[1]Radiative Energy Transfer Lab, Department of Mechanical Engineering, University of Utah, Salt Lake City, UT 84112, USA

[2]Université de Lyon, CNRS, INSA-Lyon, UCBL, CETHIL, UMR5008, F-69621 Villeurbanne, France

The impacts of radiative, electrical and thermal losses on the power output enhancement of nanoscale-gap thermophotovoltaic (nano-TPV) power generators consisting of a gallium antimonide cell paired with a broadband tungsten and a radiatively-optimized Drude radiator are analyzed. Results reveal that surface mode mediated nano-TPV power generation with the Drude radiator outperforms the tungsten emitter, dominated by frustrated modes, only for a vacuum gap thickness of 10 nm and if both electrical and thermal losses are neglected. The key limiting factors for the Drude and tungsten-based devices are respectively the recombination of electron-hole pairs at the cell surface and thermalization of radiation with energy larger than the absorption bandgap. In a nano-TPV power generator cooled by convection with a fluid at 293 K and a heat transfer coefficient of $10^4$ Wm$^{-2}$K$^{-1}$, power output enhancements of 4.69 and 1.89 are obtained for the tungsten and Drude radiators, respectively, when a realistic vacuum gap thickness of 100 nm is considered. A design guideline is also proposed where a high energy cutoff above which radiation has a net negative effect on nano-TPV power output is determined. This work demonstrates that design and optimization of nano-TPV devices must account for radiative, electrical and thermal losses.

a) Electronic mail: michael.bernardi@utah.edu
b) Electronic mail: rodolphe.vaillon@insa-lyon.fr
c) Electronic mail: mfrancoeur@mech.utah.edu

Radiation heat transfer in the near field, where bodies are separated by sub-wavelength gaps, exceeds Planck's blackbody distribution due to energy transport by tunneling of evanescent modes.[1-8] These modes include evanescent waves generated by total internal reflection of propagating waves at the material-gap interface (frustrated modes) and surface waves such as surface plasmon-polaritons[9] and surface phonon-polaritons (surface modes).[10] Whale and Cravalho[11,12] suggested that direct thermal-to-electrical energy conversion via thermophotovoltaic (TPV) power generators could benefit from the near-field effects of thermal radiation by spacing the radiator and the cell by a sub-wavelength vacuum gap. Since then, a few groups analyzed nanoscale-gap TPV (nano-TPV) power generators theoretically[13-21] and experimentally.[22-27]

Numerical studies predicted a potential power output enhancement by a factor of 20 to 30 in nano-TPV systems, but most of these modeling efforts only accounted for radiative losses in the cell.[13-18] Radiative and electrical losses in nano-TPV power generators were considered for the first time by Park et al.[19] A device consisting of a tungsten radiator at 2000 K and an indium gallium antimonide ($In_{0.18}Ga_{0.82}Sb$) cell maintained at 300 K was modeled. Results revealed that electrical losses induce a drop in conversion efficiency by 5 to 10%. The same configuration was analyzed by Bright et al.,[20] except that a gold reflecting layer on the backside of the cell was considered for recycling radiation that does not contribute to photocurrent generation. To date, only Francoeur et al.[21] considered the coupled effects of radiative, electrical and thermal losses on the performances of a nano-TPV system made of a tungsten radiator maintained at 2000 K and an $In_{0.18}Ga_{0.82}Sb$ cell. Results showed that the broadband enhancement of the radiative flux in the near field does not automatically lead to improved performance due to large thermal losses and the associated temperature increase in the cell inducing a significant drop of the power output. It was found that a thermal management system with a high heat transfer coefficient of $10^5$ Wm$^{-2}$K$^{-1}$ was required to maintain the cell at room temperature for nanometer-size gaps in order to obtain performances similar to those of Park et al. It is thus clear that accounting for the three loss mechanisms is a critical component of the design of optimal nano-TPV power generators. Indeed, Dupré and Vaillon[28] proposed a novel approach for optimizing the performance of standard crystalline silicon solar cells by minimizing radiative and electrical losses as well as thermal losses that are usually omitted. It was shown that the cell architecture leading to a maximum power output is affected when thermal losses are considered in the optimization procedure. This is expected to have an even more significant impact in the optimization of nano-TPV power generators.

Nano-TPV power generators constitute an attractive alternative to conventional TPV systems limited by the Planck blackbody distribution. Experimental nano-TPV devices are however challenging to fabricate, since a nanosize vacuum gap needs to be maintained between two surfaces having dimensions of a few hundreds of micrometers to a few millimeters. As such, this technology will be viable only if a significant power output enhancement over conventional TPV devices can be obtained. The objective of this paper is therefore to investigate in depth the impacts of radiative, electrical and thermal losses on nano-TPV power output enhancement. In particular, the contributions of propagating, frustrated and surface modes to radiative, electrical and thermal losses are analyzed in detail for nano-TPV systems with tungsten and radiatively-optimized Drude radiators. A secondary objective is to provide general guidelines for the design and conception of nano-TPV devices when all loss mechanisms are taken into account.

**Results**

**Interplay between radiative, electrical and thermal losses.** Figure 1 shows how the coupled effects of radiative, electrical and thermal losses negatively affect nano-TPV power output. Note that losses are defined relative to the power absorbed by the cell. Reflection by the cell is not a radiative loss for TPV systems, as reflected radiation can be absorbed by the radiator. Yet, reflection should be minimized in order to maximize radiation absorption by the cell. Additionally, transmission is negligible for a micrometer-thick cell. Radiation absorbed by the cell with energy $E$ below its bandgap $E_g$ does not generate electron-hole pairs (EHPs) and is classified as a radiative loss. Additionally, since a fraction of this energy is dissipated as heat via absorption by the lattice and the free carriers, it also contributes to thermal losses resulting in an increase of the cell temperature $T_{cell}$. As $T_{cell}$ increases, the dark current increases thereby decreasing the power output.[21] The radiative properties and the absorption bandgap of the cell are temperature-dependent, such that there is a feedback component, shown by the dashed arrow 1 in Fig. 1, affecting the absorption characteristics and therefore the radiative losses. Radiation absorbed by the cell with energy $E$ equal to or larger than $E_g$ generates EHPs. Electrical losses arise when the photogenerated EHPs recombine before being separated at the depletion region of the cell, thus reducing the power output. Electrical losses include recombination within the volume and at the surfaces of the cell. These mechanisms also contribute to thermal losses since the EHPs that undergo non-radiative recombination dissipate their energy as heat. As the electrical properties of the cell are temperature-dependent, an increase in $T_{cell}$ also affects recombination of EHPs; this coupling is represented by the dashed arrow 2 in Fig. 1. Radiation with energy $E$ larger than the bandgap $E_g$ dissipates its excess energy as heat through thermalization, thus contributing to thermal losses. There is a feedback component to this loss mechanism, shown by the dashed arrow 3 in Fig. 1, as increasing $T_{cell}$ lowers the absorption bandgap of the cell and modifies its radiative properties. Clearly, accounting for thermal losses substantially increases the complexity of the problem as the loss mechanisms are strongly coupled to each other.

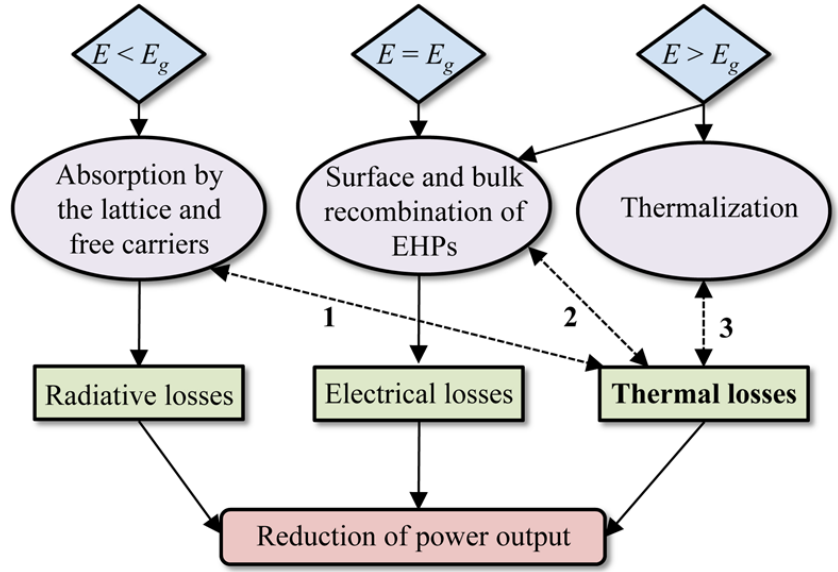

Figure 1. Coupled effects of radiative, electrical and thermal losses on the power output of nano-TPV power generators.

**Description of the problem.** The nano-TPV power generator under study is shown in Fig. 2 and consists of a semi-infinite radiator (layer 0) at a temperature of 2000 K and a 10.4-µm-thick cell (layers 2 and 3) separated by a vacuum gap of thickness $d$ (layer 1). The problem is assumed to be one-dimensional, for which only the variations along the $z$-axis normal to the surface of the radiator and the cell are accounted. The cell is modeled as a single p-n junction made of gallium antimonide (GaSb) that has a bandgap of 0.72 eV at 293 K. The thickness and doping level of the p-region (layer 2) are 0.4 µm and $10^{19}$ cm$^{-3}$, respectively, while the thickness of the n-region (layer 3) is 10 µm with a doping level of $10^{17}$ cm$^{-3}$. For these conditions, the thickness of the depletion region, assumed to be exclusively in the n-doped region, is 113 nm at 293 K.[21] A convective boundary condition with $T_\infty$ = 293 K and a heat transfer coefficient fixed at $h_\infty = 10^4$ Wm$^{-2}$K$^{-1}$ is used as a thermal management system (layer 4).

Vacuum gap thicknesses $d$ ranging from 10 to 1000 nm are considered in order to maximize radiative heat transfer by evanescent modes. In practice, maintaining a vacuum gap on the order of a few tens of nanometers between two millimeter-size surfaces is difficult to achieve. However, in the future, the bottlenecks associated with maintaining a nanosize gap may be overcome. As such, the analysis presented here is not limited to current technological constraints. Additionally, measurement of radiation heat transfer between a microsize sphere and a surface separated by a 30-nm-thick vacuum gap was reported by Rousseau et al.[5] The experimental results were compared against numerical predictions based on the Derjaguin approximation, where the flux between the sphere and the surface is computed as a summation of local heat fluxes between two parallel surfaces. The results presented hereafter could thus be used for designing a sphere-surface nano-TPV experimental bench where it is possible to maintain a nanosize gap.

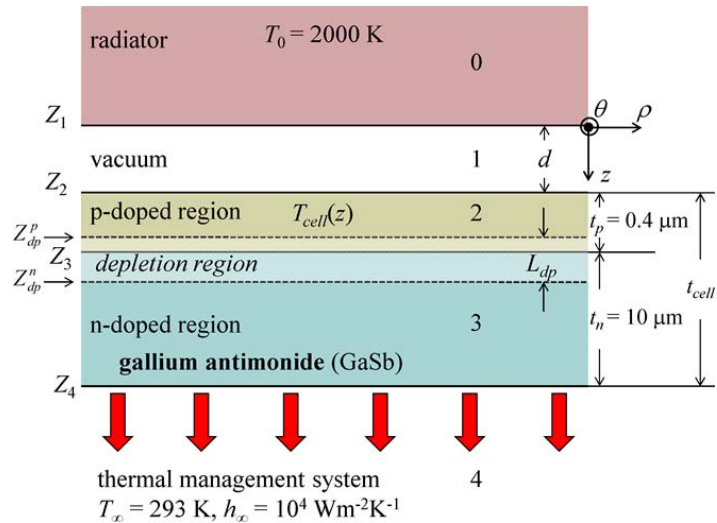

Figure 2. Nano-TPV power generator consisting of a radiator, maintained at a constant and uniform temperature of 2000 K, separated by a vacuum gap of thickness $d$ from a GaSb cell.

Various numerical studies that accounted solely for radiative losses suggested that radiators with quasi-monochromatic emission in the near field matching the bandgap of the cell result in better performing nano-TPV power generators. In order to verify this hypothesis, a radiatively-optimized radiator supporting surface polariton modes made of a fictitious material with a dielectric function described by a Drude model given by $\varepsilon(\omega) = 1 - \omega_p^2/(\omega^2 + i\Gamma\omega)$ is considered, where the plasma frequency $\omega_p$ and the loss coefficient $\Gamma$ are fixed at $1.83 \times 10^{15}$ rad/s (1.20 eV) and $2.10 \times 10^{13}$ rad/s (0.0138 eV), respectively. These values were chosen following the technique proposed by Ilic et al.[15] where $\Gamma$ is calculated to maximize radiation transfer with energy larger than the cell absorption bandgap for a vacuum gap of 10 nm. Optimizing $\Gamma$ for every gap thickness is unnecessary as it leads to a variation in the power output of less than 1%. The plasma frequency $\omega_p$ was chosen so that surface polariton resonance occurs at a radiation energy slightly above the cell bandgap. Surface polariton resonance for the radiator-vacuum interface is $\omega_p/\sqrt{2}$, which corresponds to a frequency of $1.29 \times 10^{15}$ rad/s (0.850 eV).[1] For the specific values of $\omega_p$ and $\Gamma$ selected here, the real part of the refractive index of the Drude radiator is between zero and one within the spectral band of interest (0.09 eV $\leq E \leq$ 2.5 eV), such that no frustrated modes are generated at the radiator-vacuum interface. For comparison, a tungsten radiator where the emission in the near field is dominated by frustrated modes is also considered. The dielectric function of tungsten has been obtained by curve-fitting the data provided in Ref. 29. Note that despite supporting surface plasmon-polaritons in the near infrared, tungsten does not exhibit quasi-monochromatic near-field thermal emission due to high losses. For the tungsten-vacuum interface, resonance occurs when the real part of the permittivity $\varepsilon' = -1$ which corresponds to a frequency of $1.97 \times 10^{15}$ rad/s (1.30 eV). At this frequency, the imaginary part of the permittivity $\varepsilon''$ is large and takes a value of 20.3.

In the simulations, the cell is discretized into $N$ control volumes and its temperature is initialized at 293 K. The radiative energy absorbed in each control volume is calculated from fluctuational electrodynamics,[30] and is used to determine the net radiation absorbed by the cell due to the lattice and the free carriers, the heat losses by thermalization and the local generation rate of EHPs. The generated photocurrent and heat sources due to non-radiative and radiative recombination are afterwards calculated by solving the minority carrier diffusion equations. Note that radiative recombination has a negligible effect on the overall energy balance of the cell.[21] An updated temperature of the cell is obtained by solving the energy equation. The radiative, electrical and thermophysical properties of the GaSb cell, provided in Ref. 21, are calculated at the updated cell temperature, and computations are repeated until $T_{cell}$ converges. The dark current is obtained by solving the minority carrier diffusion equations without the local generation rate of EHPs, and various performance indicators, such as the power output, are finally calculated. The details of the computational model are provided in the Methods section.

**Impacts of radiative, electrical and thermal losses on nano-TPV power output enhancement.** Figure 3 shows the power output enhancement of the tungsten- and radiatively-optimized Drude-based nano-TPV power generators as a function of the vacuum gap thickness $d$ and the type of losses considered in the model. For both radiators, the power output enhancement is defined as the power output of the actual device ($P$) over the power output obtained with a tungsten source in the far field ($P_{FF} = 3.18 \times 10^4$ Wm$^{-2}$, $T_{cell} = 298$ K) when all loss mechanisms are considered. Note that when thermal losses are neglected, the temperature of the cell is fixed at 293 K.

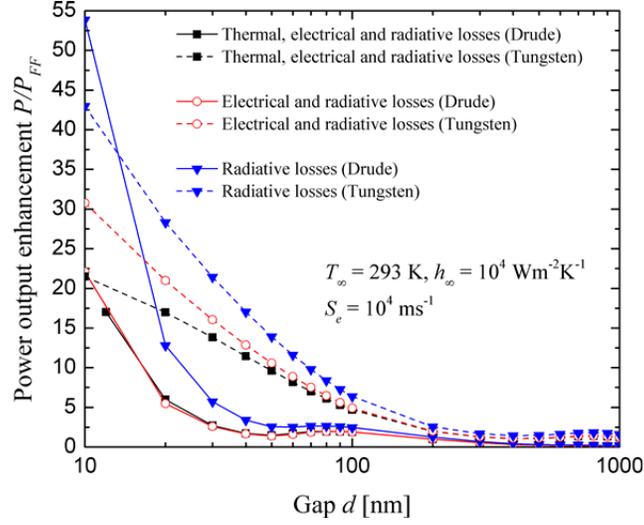

Figure 3. Power output enhancement as a function of the vacuum gap thickness $d$ and the type of losses considered for tungsten and Drude radiators.

The power output enhancement increases as the thickness of the vacuum gap decreases due to an increasing contribution of evanescent modes to the radiative flux. The only exception arises for the Drude radiator, where the power output enhancement increases non-monotonically for sub-100 nm vacuum gap thicknesses; the local minimum observed at $d = 50$ nm will be explained when analyzing the separate contributions of propagating, frustrated and surface modes. For a fixed gap thickness and for both radiators, the power output enhancement $P/P_{FF}$ is maximum when only radiative losses are considered, where it is assumed that all radiation absorbed by the cell with energy $E$ larger than the bandgap $E_g$ generates EHPs contributing to the photocurrent (i.e., the quantum efficiency is 100% for radiation with $E \geq E_g$). When radiative and electrical losses are taken into account, the quantum efficiency for radiation with $E \geq E_g$ is no longer 100% due to bulk and surface recombination of EHPs thus resulting in a lower power output enhancement. For the case that radiative, electrical and thermal losses are considered, heat generation in the cell due to absorption by the lattice and the free carriers, thermalization and recombination of EHPs increases the temperature of the cell, thus decreasing the power output enhancement because of an increase of the dark current.

The maximum power output enhancement, arising for a 10-nm-thick gap, is 43.0 and 53.9 for the tungsten- and the Drude-based devices, respectively, when only radiative losses are considered. When both electrical and thermal losses are added, the temperature of the cell and the power output enhancement for the tungsten radiator are respectively 448 K and 21.5 (-50.0% relative to radiative losses), while the cell exceeds its melting temperature of 985 K for the Drude-based system. The equilibrium temperatures of both types of nano-TPV devices as a function of the vacuum gap thickness are provided in Fig. S.1 of the Supplemental Information section. Note that in all simulations, the temperature difference in the cell is negligible ($\Delta T_{cell} \leq 1.3$ K), such that a single average temperature is reported for a given vacuum gap thickness. It is also interesting to note that the radiatively-optimized Drude radiator leads to a power output enhancement larger than that achieved with tungsten only for the case $d = 10$ nm when only radiative losses are accounted for.

Figure 4 shows the power output enhancement as a function of the vacuum gap thickness, the type of losses considered and the modes contributing to power generation. These modes are defined relative to the radiator-vacuum interface. Propagating modes propagate in both the radiator and the vacuum gap, and correspond to waves with parallel wavevectors $k_\rho$ smaller than $k_0$, where $k_0$ is the magnitude of the wavevector in vacuum. Frustrated modes are propagating within the radiator and evanescent in the vacuum gap, such that they are described by parallel wavevectors $k_0 < k_\rho < n'k_0$, where $n'$ is the real part of the refractive index of the radiator. Finally, surface polariton modes are evanescent on both sides of the radiator-vacuum interface and are described by parallel wavevectors $k_\rho > n'k_0$.

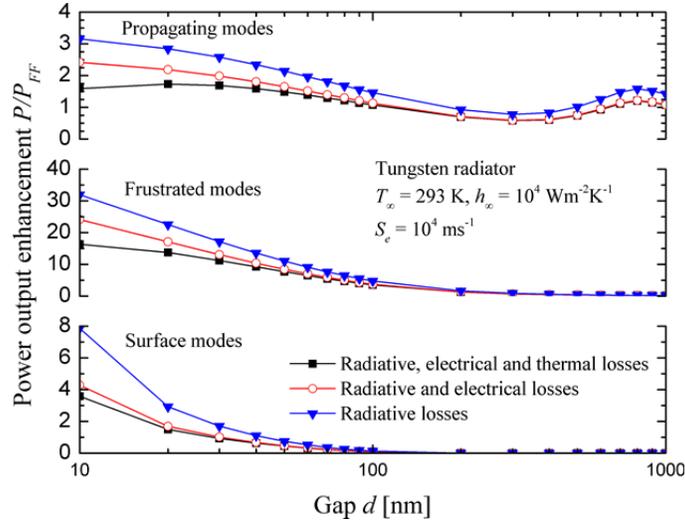

(a)

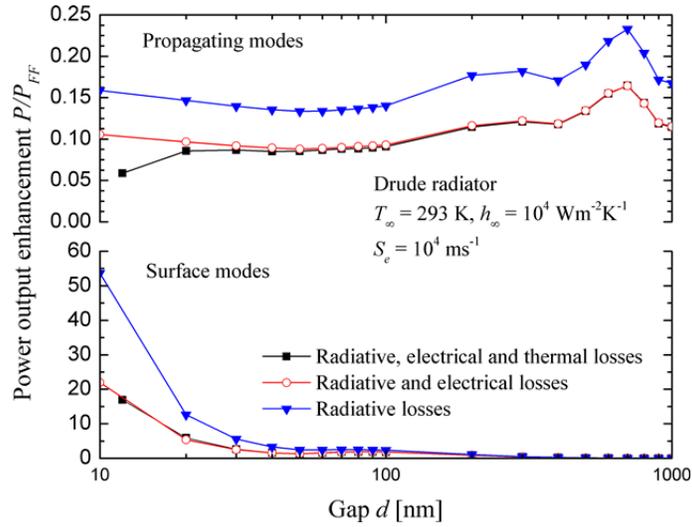

(b)

Figure 4. Power output enhancement as a function of the vacuum gap thickness $d$ and the type of losses considered showing the contributions of propagating, frustrated and surface modes: (a) tungsten radiator. (b) Drude radiator.

For the tungsten radiator shown in Fig. 4(a), the power output enhancement in the near field is dominated by frustrated modes. For example, at $d = 10$ nm, the relative contributions of propagating, frustrated and surface modes to $P/P_{FF}$ are respectively 7.4, 75.9 and 16.7% when all loss mechanisms are considered. Note that that the power output enhancement due to propagating waves exceeds unity even when radiative, electrical and thermal losses are included in the model. In the far field, the cell absorbs a propagating flux of $1.05 \times 10^5$ Wm$^{-2}$ and this number increases to $2.24 \times 10^5$ Wm$^{-2}$ when the gap thickness is 10 nm. The increased absorption of propagating waves in the near field is due to the decrease of the absorption bandgap of the cell as $T_{cell}$ increases. When the tungsten radiator is in the far field, the cell temperature reaches 298 K and the absorption bandgap is 0.72 eV. For a gap thickness of 10 nm, the cell temperature reaches 448 K leading to a lower absorption bandgap of 0.66 eV. Additionally, propagating wave interference could also explain the local minimum in power output enhancement observed at a gap thickness around 300 nm.[31]

For the Drude radiator shown in Fig. 4(b), the power output enhancement is dominated by surface modes (no frustrated modes are generated). The relative contribution of surface modes to power generation is more than 99% for gap thicknesses equal to or smaller than 20 nm regardless of the types of losses considered. Figure 4(b) also shows that the local minimum observed for a 50-nm-thick gap is due to surface modes. This minimum arises due to a modification of the dispersion relation of surface modes supported at the radiator-vacuum interface caused by the presence of the cell. At gap thicknesses equal to or smaller than 30 nm, surface polariton resonance at an angular frequency of $\omega_p / \sqrt{2}$ (0.85 eV) dominates energy transfer. At gap thicknesses equal to or larger than 80 nm, radiative energy transfer is dominated by a second peak at a lower radiation energy (~ 0.75 eV). As outlined by Laroche et al.,[14] this peak corresponds to the non-asymptotic portion of the dispersion relation of surface modes that is broadened by the cell. As the broadening of the dispersion relation varies as a function of the vacuum gap thickness, the spectral location and strength of the second peak varies with gap thickness. The local minimum at $d = 50$ nm is due to the transition between the two peaks, as shown in Fig. S.2 of the Supplemental Information section.

The Drude radiator used here has been optimized by minimizing radiative losses. It is clear however that even if only radiative losses are considered, nano-TPV power generation dominated by surface modes is useful in the extreme near field for a vacuum gap thickness on the order of 10 nm. Indeed, the radiative flux absorbed by the cell with the tungsten radiator decreases at a slower rate than with the Drude radiator (see Fig. S.3 of the Supplemental Information section). For vacuum gap thicknesses from 10 to 50 nm, the total radiative heat flux absorbed with the Drude radiator decays at a rate proportional to $d^{-1.82}$ when only radiative losses are considered. Note that the fact that radiative energy is being transferred between dissimilar materials results in a slight deviation from the $d^{-2}$ power law observed for surface polariton mediated energy transfer between identical materials.[32]

Figure 4 suggests that the contribution of surface modes to the power output is more affected by electrical losses than the contribution by frustrated modes. Surface polaritons are modes with large momentum (i.e., large parallel wavevector $k_\rho$) and thus small penetration depth on the order of the vacuum gap separating the radiator and the cell.[32,33] On the other hand, the penetration depth of frustrated modes is on the order of the wavelength in the medium and is independent of the vacuum gap thickness. Consequently, for nanosize gaps, recombination of EHPs at, or near, the surface of the cell is likely to be a limiting factor to surface polariton mediated nano-TPV power generation. The diffusion length of the minority carrier electrons in the p-region of the cell can be estimated as $L_e = \sqrt{D_e \tau_e}$. For GaSb at 293 K, the electron diffusion coefficient $D_e$ and lifetime $\tau_e$ are 29.1 cm²s⁻¹ and 5.70 ns,[21] respectively, thus resulting in a diffusion length $L_e$ of 4.07 μm. Therefore, bulk recombination of EHPs is not the limiting factor to surface polariton mediated nano-TPV power generation since $L_e$ is much larger than the thickness of the p-doped region. A surface recombination velocity $S_e$ of $10^4$ ms$^{-1}$ at the top surface of the cell was used for generating the results shown in Figs. 3 and 4, which is a typical value for GaSb.[34] In order to further investigate the effect of surface recombination on nano-TPV performance, simulations have been performed using a surface recombination velocity of $S_e = 0$ ms$^{-1}$ (see Fig. S.4 of the Supplemental Information section). Although completely eliminating surface recombination is unrealistic, it is possible to minimize $S_e$ by decreasing the surface roughness and/or passivating the surface, at least in silicon-based photovoltaics.[35-39] For a 10-nm-thick gap, the tungsten-based system exhibits a power output enhancement of 41.6, a 3.3% relative decrease when compared to the case where only radiative losses are considered and a relative increase of 88.2% over the case that includes electrical losses with surface recombination. For the same gap thickness, the Drude-based system has a power output enhancement of 53.6. This is a 0.5% relative decrease from the case with radiative losses only and a 143% relative increase over the case that includes electrical losses with surface recombination. These results show that surface recombination is a limiting factor to nano-TPV power generation due to the small penetration depth associated with evanescent modes. Surface recombination has a larger effect on the performance of the Drude-based system than the tungsten-based device due to the fact that surface polariton modes are characterized by larger parallel wavevectors than frustrated modes and thus smaller penetration depths in the cell. Surface polariton mediated nano-TPV power generation is thus potentially interesting at very small gap thicknesses and when surface recombination velocity is minimized, although those conditions may be difficult to achieve in practice. Alternatively, surface polariton modes may perform better with thin cells on the order of a few tens of nanometers.

Despite a large heat transfer coefficient of $10^4$ Wm$^{-2}$K$^{-1}$ for the thermal management system, thermal losses have a significant negative impact on nano-TPV power generation. For the Drude radiator, Figs. 3 and 4(b) show that thermal losses have a negligible effect on $P/P_{FF}$, when compared to the case of radiative and electrical losses, down to a gap thickness of 20 nm. The cell reaches a temperature of 350 K for $d$ = 20 nm, as opposed to 388 K when the tungsten radiator is used (see Fig. S.1). On the other hand, the temperature of the cell rapidly increases with the radiatively-optimized Drude radiator for gap thicknesses below 20 nm and eventually exceeds its melting point slightly below a 12-nm-thick gap. It can also be observed in Figs. 3 and 4(b) that in the gap range from 20 to 80 nm where $T_{cell}$ varies from 350 to 303 K, respectively, the power output enhancement slightly exceeds the predictions at 293 K where radiative and electrical losses are considered. This counterintuitive behavior is explained by the fact that the bandgap decreases as $T_{cell}$ increases, thus increasing radiation absorption and EHP generation. However, above a certain temperature, thermal effects become dominant and cause the power output to deteriorate rapidly due to a large dark current.

The relative contributions of thermalization, bulk non-radiative recombination of EHPs, surface recombination of EHPs, and absorption by the lattice and the free carriers due to propagating, frustrated and surface modes to heat generation $Q$ are plotted in Fig. 5 as a function of the gap thickness $d$. The units for heat generation due to surface recombination ($Q_{SR}$) are Wm$^{-2}$, while the remaining contributions within the cell have units of Wm$^{-3}$. For comparison purposes, thermalization ($Q_T$), bulk non-radiative recombination ($Q_{NRR}$) and absorption by the lattice and the free carriers ($Q_{LFC}$) are integrated over the thickness of the cell in order to obtain units of Wm$^{-2}$. Note that the magnitude of the total heat generation ranges from 5.43×10$^4$ Wm$^{-2}$ at $d$ = 1000 nm to 1.54×10$^6$ Wm$^{-2}$ at $d$ = 10 nm for the tungsten-based system; for the Drude radiator, the total heat generation varies from 1.41×10$^4$ Wm$^{-2}$ at $d$ = 1000 nm to 1.99×10$^6$ Wm$^{-2}$ at $d$ = 12 nm.

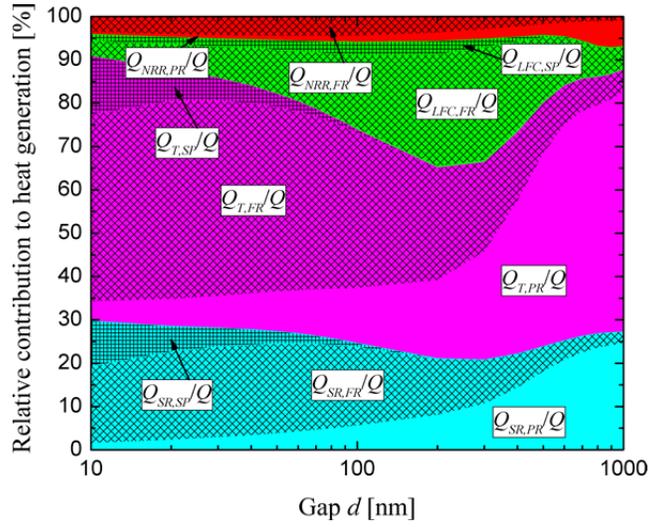

(a)

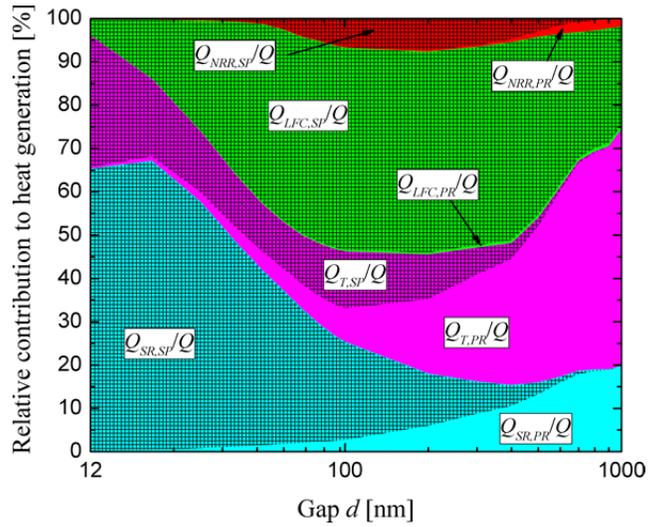

(b)

Figure 5. Relative contributions of thermalization (*T*), bulk non-radiative recombination (*NRR*), surface recombination (*SR*) and absorption by the lattice and the free carriers (*LFC*) due to propagating (*PR*), frustrated (*FR*)

and surface (*SP*) modes to heat generation within the cell as a function of the vacuum gap thickness *d*: (a) Tungsten radiator (note that negligible contributions to the total heat generation are not shown). (b) Drude radiator.

For the tungsten radiator (Fig. 5(a)), thermal losses are dominated by thermalization of frustrated modes for sub-300 nm gap thicknesses and by thermalization of propagating modes for larger gaps. This is explained by the fact that thermal emission by a tungsten radiator is broadband such that the near field enhancement occurs at all frequencies. A large portion of the radiation absorbed by the cell thus contributes simultaneously to EHP generation and thermalization. After some limiting photon energy $E$ ($> E_g$), it is reasonable to expect that radiation absorption has a net negative effect on the cell arising when the reduction of the power output due to heat generation overcomes the power produced by EHP photogeneration. This will be analyzed further in the next sub-section.

For the Drude radiator (Fig. 5(b)), thermal losses at small gap distances are dominated by surface recombination of EHPs generated by surface modes. It is therefore imperative to minimize the surface recombination velocity when radiation transfer is dominated by surface modes in order to minimize electrical and thermal losses. However, one must keep in mind that even if surface recombination is minimized, surface polariton mediated nano-TPV power generation outperforms traditional radiators only in the extreme near field. Thermalization is not as critical for the Drude radiator when compared to tungsten, since radiative heat transfer in the near field is quasi-monochromatic. The relative contribution of thermalization increases as *d* increases, when the flux contains a non-negligible portion of broadband propagating waves.

From the results obtained here, it is obvious that a Drude radiator optimized by accounting solely for radiative losses is not a viable solution for enhancing power generation in nano-TPV devices.

The design of nano-TPV systems maximizing power generation must account for all three loss mechanisms, radiative, electrical and thermal.

**Estimation of cut-off spectral band for improved nano-TPV performance.** The results discussed in the previous section suggest that nano-TPV systems exploiting frustrated modes outperform devices capitalizing on surface modes, except in the extreme near field when both electrical and thermal losses are minimized. However, broadband radiators supporting frustrated modes lead to high thermal losses due to thermalization. This section attempts to improve the performance of a tungsten-based device ($d$ = 10 nm) dominated by frustrated modes by filtering a portion of the near-field emission spectrum while accounting for all loss mechanisms.

Radiation with energy $E$ lower than the absorption bandgap of the cell $E_g$ should be suppressed as it only contributes to radiative and thermal losses (see Fig. 1). For simplicity, a fixed low energy cutoff of 0.66 eV is selected since the cell temperature is expected to be higher than 293 K due to thermal losses. This choice is motivated by the fact that when all losses are considered, the device with the tungsten radiator and $d$ = 10 nm reaches an equilibrium temperature of 448 K; at this temperature, the resulting absorption bandgap of the cell is 0.66 eV. In addition, the reduction of the power output associated with heat dissipation in the cell due to absorption of radiation with energy $E$ larger than $E_g$, dominated by thermalization, may exceed the power produced from the generation of EHPs. As such, a high energy cutoff $E_{high}$ above which radiation absorption has a net negative effect on the power output can be determined. This is done by analyzing the difference between the power output obtained with and without filter ($\Delta P = P_{filter} - P$, where $P = 6.82 \times 10^5$ W/m$^2$) as a function of $E_{high}$ (see Fig. 6).

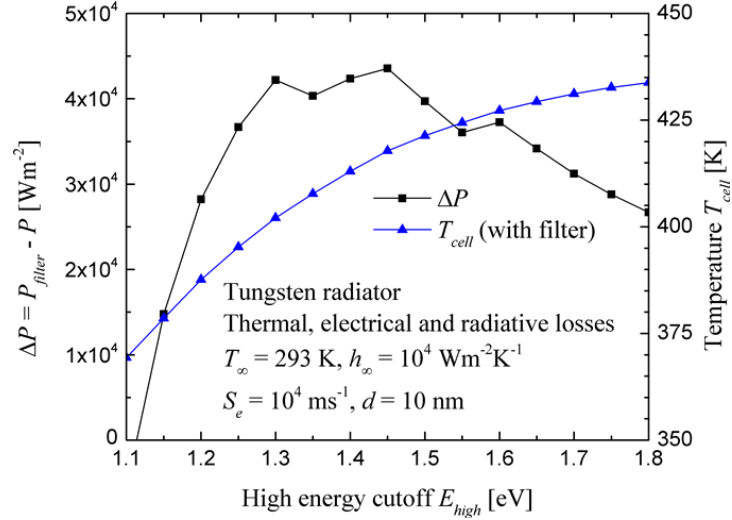

Figure 6. Difference of power output with and without filter $\Delta P$ and cell temperature $T_{cell}$ as a function of the high energy cutoff $E_{high}$ (the low energy cutoff is fixed at 0.66 eV).

There is an optimum high energy cutoff $E_{high}$ (1.45 eV) at which the difference in power output $\Delta P$ is maximized. This arises since the cell temperature decreases with decreasing the high energy cutoff due to smaller heat losses by thermalization (Fig. 5(a)). As such, the maximum voltage increases while the maximum current decreases since a smaller number of EHPs are generated by decreasing the high energy cutoff (see Fig. S5 in the Supplemental Information section). By limiting thermal emission by the tungsten source to a spectral band from 0.66 to 1.45 eV, a power output enhancement of 22.9 is achieved. This is a 6.5% increase compared to the power output enhancement when considering the full spectrum of tungsten, where a $P/P_{FF}$ value of 21.5 was obtained. It is also remarkable that the cell reaches a temperature of 418 K, while a temperature of 448 K was obtained with the entire spectrum. By analyzing thermal losses that are usually neglected in the literature, it is possible to decrease the cell temperature by 30 K (and thus reduce the cooling load) and to increase the power output enhancement by 6.5%.

It is clear that thermal losses are an integral part of the design of optimal nano-TPV power generators. Here, it was possible to determine a spectral limit above which radiation absorption is more detrimental than beneficial to the power output by accounting for heat generation within the cell. A rigorous optimization scheme would require further analysis for determining the near-field thermal spectrum maximizing power generation.

## Conclusions

This work demonstrated that a quasi-monochromatic radiator supporting surface modes outperforms a tungsten radiator only in the extreme near field when both electrical and thermal losses are minimized. In an actual nano-TPV power generator and for a realistic vacuum gap thickness of 100 nm, which would automatically account for all loss mechanisms, a power output enhancement of 4.69 is achieved through the use of a tungsten radiator, while a power output enhancement factor of 1.89 is obtained with a radiatively-optimized Drude radiator. Recombination of EHPs at the surface of the cell is the main limiting factor to surface mode mediated nano-TPV power generation. There are well-documented methods for reducing the surface recombination velocity of silicon photovoltaic cells but it may still be an open topic for lower absorption bandgap TPV cells. For a radiator dominated by frustrated modes, thermalization significantly affects nano-TPV power output due to the broadband enhancement of the flux in the near field. Finally, a preliminary analysis allowed the determination of a high energy cutoff above which radiation absorption has a net negative effect on nano-TPV power output. Such an analysis is only possible when thermal losses are considered, thus showing that radiative, electrical and thermal losses must be considered in the design of optimal nano-TPV power generators.

## Methods

The net radiative heat flux absorbed by a control volume of thickness $\Delta z_j$ in the cell is calculated using fluctuational electrodynamics and the dyadic Green's functions for layered media.[21,30] The absorbed radiation with energy equal to or larger than the bandgap of the cell is used for calculating the local monochromatic generation rate of EHPs $g_\omega(z)$ [m$^{-3}$s$^{-1}$(rad/s)$^{-1}$], acting as a source term in the minority carrier diffusion equations:[21,40]

$$D_e \frac{d^2 \Delta n_{e,\omega}(z)}{dz^2} - \frac{\Delta n_{e,\omega}(z)}{\tau_e} + g_\omega(z) = 0 \tag{1a}$$

$$D_h \frac{d^2 \Delta n_{h,\omega}(z)}{dz^2} - \frac{\Delta n_{h,\omega}(z)}{\tau_h} + g_\omega(z) = 0 \tag{1b}$$

where the dependent variables $\Delta n_{e,\omega}$ and $\Delta n_{h,\omega}$ are the local excess of minority carriers (*e*: electrons in the p-region, *h*: holes in the n-region) above the equilibrium concentration [m$^{-3}$], $D_e$ and $D_h$ are the minority carrier diffusion coefficients [m$^2$s$^{-1}$], while $\tau_e$ and $\tau_h$ are the minority carrier lifetimes [s] that include radiative and non-radiative (Auger and Shockley-Read-Hall) recombination processes. The local monochromatic generation rates of EHPs within a given control volume *j* in the p- and n-regions are respectively given by:

$$g_{j,\omega} = \frac{1}{\hbar\omega} \kappa_{j,\omega}^{ib} \left( \frac{q_{\omega,\Delta z_j^p}^{abs}}{\Delta z_j^p \kappa_{j,\omega}} \right) \tag{2a}$$

$$g_{j,\omega} = \frac{1}{\hbar\omega} \kappa_{j,\omega}^{ib} \left( \frac{q_{\omega,\Delta z_j^n}^{abs}}{\Delta z_j^n \kappa_{j,\omega}} \right) \tag{2b}$$

where $\omega$ is the angular frequency [rad/s], $\hbar$ is the reduced Planck constant [1.0546×10$^{-34}$ Js], $q_{\omega,\Delta z_j^{(p,n)}}^{abs}$ is the local monochromatic radiative heat flux absorbed by control volume *j* [Wm$^-$

$^2$(rad/s)$^{-1}$], $\kappa_{j,\omega}^{ib}$ is the local monochromatic interband absorption coefficient [m$^{-1}$] and $\kappa_{j,\omega}$ is the local monochromatic absorption coefficient that accounts for absorption by the lattice and the free carriers as well as the interband absorption process [m$^{-1}$].

The boundary conditions of the minority carrier diffusion equations at the cell-vacuum interface ($z = Z_2$) and at the cell-thermal management interface ($z = Z_4$) account for surface recombination of EHPs and are respectively given by:

$$D_e \frac{d\Delta n_{e,\omega}(Z_2)}{dz} = S_e \Delta n_{e,\omega}(Z_2) \tag{3a}$$

$$D_h \frac{d\Delta n_{h,\omega}(Z_4)}{dz} = S_h \Delta n_{h,\omega}(Z_4) \tag{3b}$$

where $S_e$ and $S_h$ are the minority carrier surface recombination velocities [ms$^{-1}$]. At the boundaries of the depletion region, it is assumed that the minority carriers are swept by the electric field at the p-n junction such that $\Delta n_{e,\omega}(Z_{dp}^p) = 0$ and $\Delta n_{h,\omega}(Z_{dp}^n) = 0$.

The photocurrent produced by the cell is the sum of contributions due to EHPs generated outside the depletion region diffusing to the boundaries of that zone and EHPs generated directly in the depletion region. In the depletion region, it is assumed that all EHPs contribute to the photocurrent:

$$J_{dp,\omega} = e \int_{Z_{dp}^p}^{Z_{dp}^n} g_{j,\omega} \, dz \tag{4}$$

where $e$ is the electron charge [$1.6022 \times 10^{-19}$ J(eV)$^{-1}$] and the monochromatic photocurrent has units of Am$^{-2}$(rad/s)$^{-1}$. The photocurrent generated at the boundaries of the depletion region is calculated using the solution of the minority carrier diffusion equations:

$$J_{e,\omega} = eD_e \frac{d\Delta n_{e,\omega}(Z_{dp}^p)}{dz} \tag{5a}$$

$$J_{h,\omega} = -eD_h \frac{d\Delta n_{h,\omega}(Z_{dp}^n)}{dz} \tag{5b}$$

The total photocurrent $J_{ph}$ is calculated by integrating the sum of Eqs. (4), (5a) and (5b) over the frequency from $\omega_g$ to infinity, where $\omega_g$ is the absorption bandgap of the cell in rad/s ($\omega_g = E_g e/\hbar$).

Heat transport within the cell is modeled via the one-dimensional steady-state energy equation with heat generation:

$$k\frac{d^2 T_{cell}(z)}{dz^2} + Q(z) = 0 \tag{6}$$

where $k$ is the thermal conductivity of the cell [Wm$^{-1}$K$^{-1}$]. The local volumetric heat generation [Wm$^{-3}$] is defined as:

$$Q(z) = -Q_{LFC}(z) + Q_T(z) + Q_{NRR}(z) - Q_{RR}(z) \tag{7}$$

Bulk non-radiative recombination, $Q_{NRR}$, is a heat generation process due to EHPs recombining before reaching the depletion region. EHP recombination may also result in radiation emission, $Q_{RR}$, which has a cooling effect on the cell. These two contributions are computed from the solution of the minority carrier diffusion equations.[21] The local radiative source term, $Q_{LFC}$, represents the balance between thermal emission and absorption by the lattice and the free carriers. Finally, heat dissipation within the cell by thermalization is accounted for via the term $Q_T$. These last two contributions are calculated from the solution of the near-field thermal radiation problem.[21]

At the cell boundaries, internal heat conduction and surface recombination of EHPs are balanced with an external heat flux. At the cell-vacuum interface ($z = Z_2$), the external heat flux is nil such that the boundary condition is given by:

$$k\frac{dT_{cell}(Z_2)}{dz} = S_e e E_g \int_{\omega_g}^{\infty} \Delta n_{e,\omega}(Z_2) d\omega \tag{8}$$

The boundary condition at $z = Z_4$ adjacent to the thermal management system includes the external heat flux due to convection:

$$-k\frac{dT_{cell}(Z_4)}{dz} = h_{\infty}[T_{cell}(Z_4) - T_{\infty}] \tag{9}$$

where surface recombination is neglected due to the large thickness of the cell.[41]

Solution of the energy equation provides an updated temperature of the cell. The radiative, electrical and thermophysical properties of the cell, given in Ref. 21, are then calculated at the updated cell temperature. The computations are repeated until the cell temperature converges. It has been verified that a relative convergence criterion of $10^{-4}$ on the cell temperature is sufficient and was used in all simulations.

Once convergence is reached, the minority carrier diffusion equations are solved in dark conditions ($g_{\omega}(z) = 0$) for a series of forward biases $V_f$ [V] in order to compute the dark current $J_0$ [Am$^{-2}$].[21] The photocurrent-voltage (J-V) characteristic of the nano-TPV device is then determined by calculating $J(V_f) = J_{ph} - J_0(V_f)$ as a function of the forward bias starting with $V_f = 0$. The maximum power output $P$ [Wm$^{-2}$] of the nano-TPV device is determined directly from the J-V characteristic.

**Acknowledgements**

This work was supported by the National Science Foundation under Grant No. CBET-1253577.

## Author contributions

The work was conceived by M.P.B., M.F. and R.V. The simulations were performed by M.P.B. under the supervision of M.F. with inputs from R.V., O.D., E.B. and P.-O.C. The manuscript was written by M.P.B. and M.F. with comments from all authors.

## Additional information

Competing financial interests: The authors declare no competing financial interests.

Supplemental Information: Impacts of propagating, frustrated and surface modes on radiative, electrical and thermal losses in nanoscale-gap thermophotovoltaic power generators

Michael P. Bernardi,[1,a)] Olivier Dupré,[2] Etienne Blandre,[2] Pierre-Olivier Chapuis[2], Rodolphe Vaillon,[2,b)] and Mathieu Francoeur[1,c)]

[1]Radiative Energy Transfer Lab, Department of Mechanical Engineering, University of Utah, Salt Lake City, UT 84112, USA

[2]Université de Lyon, CNRS, INSA-Lyon, UCBL, CETHIL, UMR5008, F-69621 Villeurbanne, France

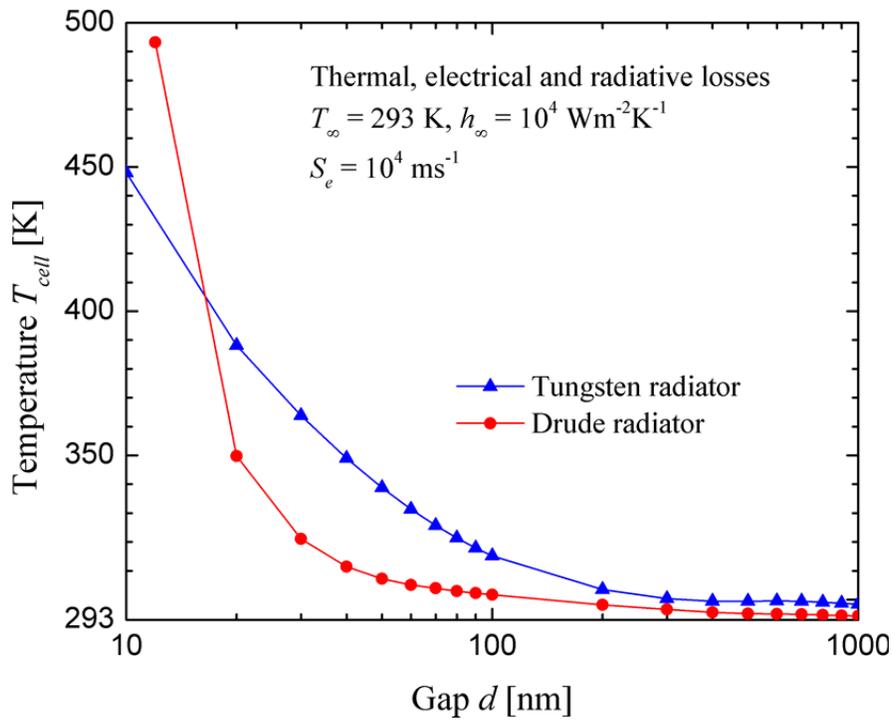

Figure S.1 Equilibrium cell temperature as a function of the vacuum gap thickness for tungsten and Drude radiators.

[a)] Electronic mail: michael.bernardi@utah.edu
[b)] Electronic mail: rodolphe.vaillon@insa-lyon.fr
[c)] Electronic mail: mfrancoeur@mech.utah.edu

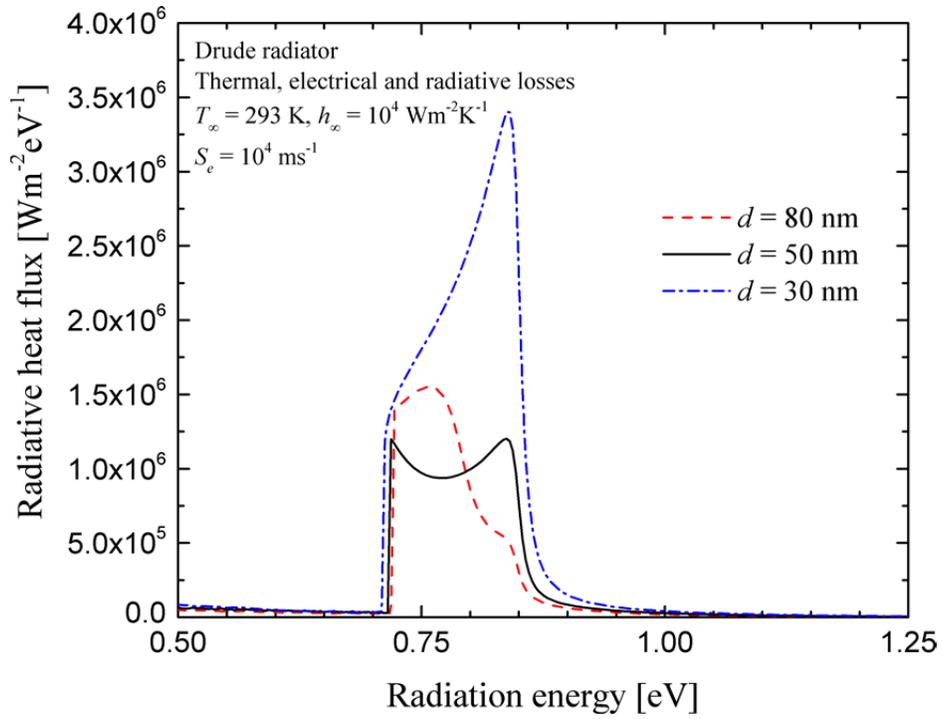

Figure S.2 Spectral distribution of radiative heat flux at the surface of the cell ($z = Z_2$) for vacuum gap thicknesses of 30, 50 and 80 nm (Drude radiator).

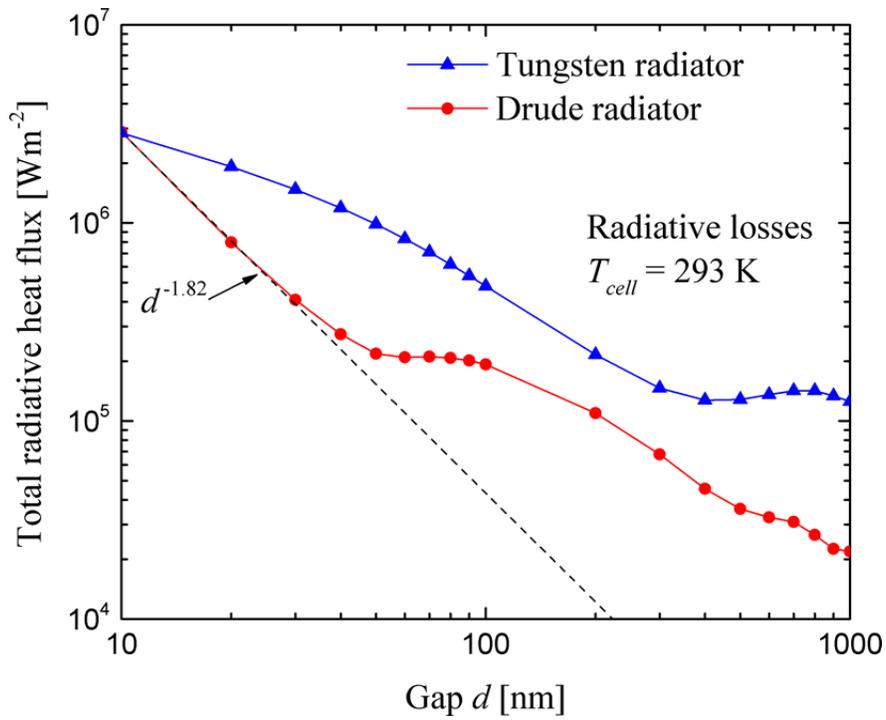

Figure S.3 Total radiative flux absorbed by the cell as a function of the vacuum gap thickness for tungsten and Drude radiators.

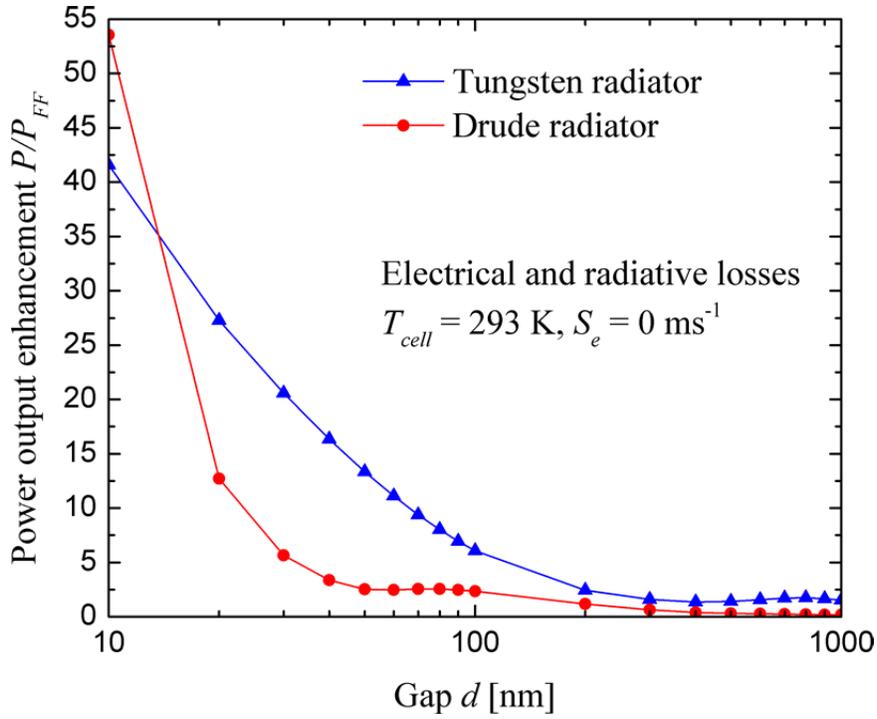

Figure S.4 Power output enhancement as a function of the vacuum gap thickness when neglecting surface recombination velocity and thermal losses for tungsten and Drude radiators.

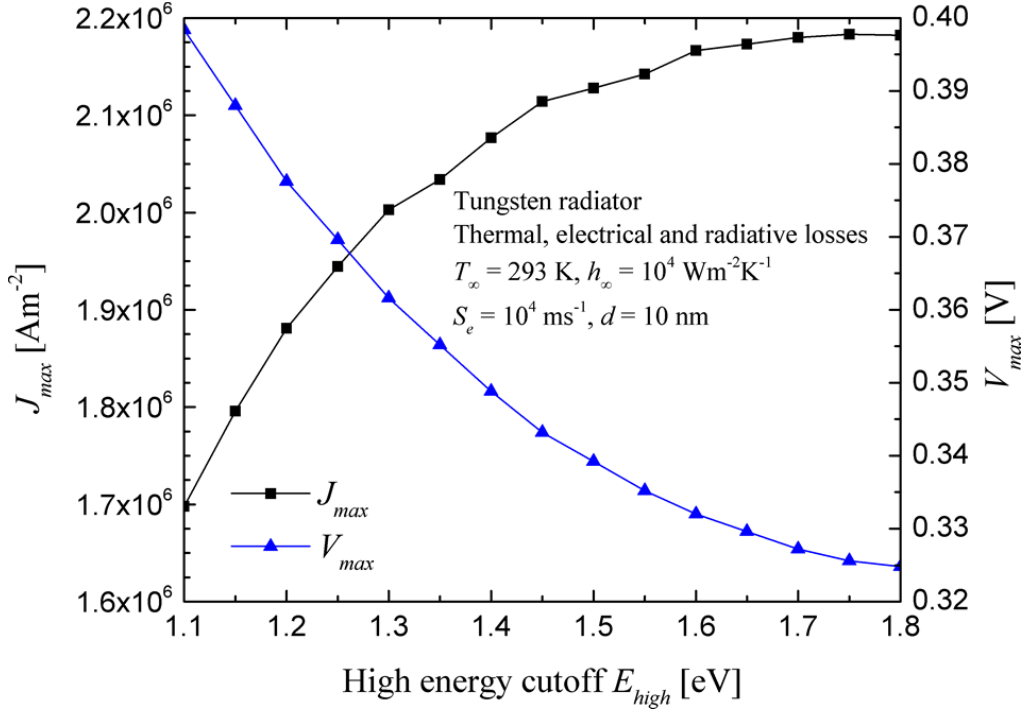

Figure S.5 Photocurrent $J_{max}$ and potential $V_{max}$ at maximum power output as a function of the high energy cutoff $E_{high}$.

Figure S.5 shows photocurrent $J_{max}$ and voltage $V_{max}$ at the maximum power output as a function of the high energy cutoff $E_{high}$. As expected, $J_{max}$ increases as $E_{high}$ increases due to a larger number of EHPs generated. Conversely, $V_{max}$ decreases as $E_{high}$ increases because of an increasing thermalization heat source and consequently a rise in temperature leading to a larger dark current.